\pdfoutput=1
\documentclass{moriond}

\usepackage{siunitx}

\def\be{\begin{equation}}
\def\ee{\end{equation}}
\def\bea{\begin{eqnarray}}
\def\eea{\end{eqnarray}}

\newcommand{\Photo}{}

\begin{document}

\vspace*{4cm}
\title{SEARCH FOR VECTOR-LIKE QUARK, HEAVY NEUTRAL LEPTON AND LONG-LIVED PARTICLES AT ATLAS AND CMS}

\author{S. GRANCAGNOLO,\\ on behalf of the ATLAS and CMS Collaborations}

\address{Dipartimento di Matematica e Fisica ``Ennio De Giorgi'',\\ Università del Salento, Via per Arnesano - 73100 Lecce (LE), Italy}

\maketitle\abstracts{
  Several theories beyond the Standard Model (BSM) predict heavy neutral leptons, or long-lived particles with unique signatures which are difficult to reconstruct. Another area of interest are vector-like quarks which lie at the heart of many extensions to the Standard Model seeking to address the Hierarchy Problem, as they can naturally cancel the mass divergence for the Higgs boson. New results for these BSM searches from the ATLAS and CMS experiments at the LHC are presented. }

\section{Introduction}
New results from the ATLAS\cite{PERF-2007-01} and CMS\cite{CMS-CMS-00-001} experiments at the Large Hadron Collider (LHC) are presented on searches for new heavy neutral leptons (HNL, Section~\ref{Section:HNL}), long-lived particles (LLP, Section~\ref{Section:LLP}), and vector-like quarks (VLQ, Section~\ref{Section:VLQ}). If existing, they could, e.g., provide a dark matter candidate, among other answers to currently unexplained phenomena. No signals of physics beyond the Standard Model (SM) were observed, and limits were therefore set on various theoretical and kinematic parameter ranges, superseding those obtained in the past.\footnote{Copyright 2024 CERN for the benefit of the ATLAS and CMS Collaborations. Reproduction of this article or parts of it is allowed as specified in the CC-BY-4.0 license.}

\section{Heavy Neutral Leptons}\label{Section:HNL}
\begin{figure}
\begin{minipage}{0.21\linewidth}
\centerline{\includegraphics[width=\linewidth]{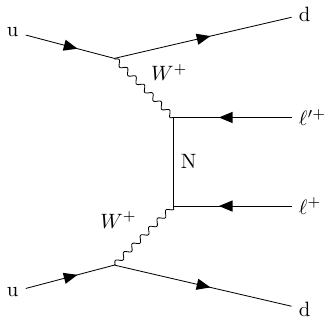}}
\end{minipage}
\hfill
\begin{minipage}{0.42\linewidth}
\centerline{\includegraphics[width=\linewidth]{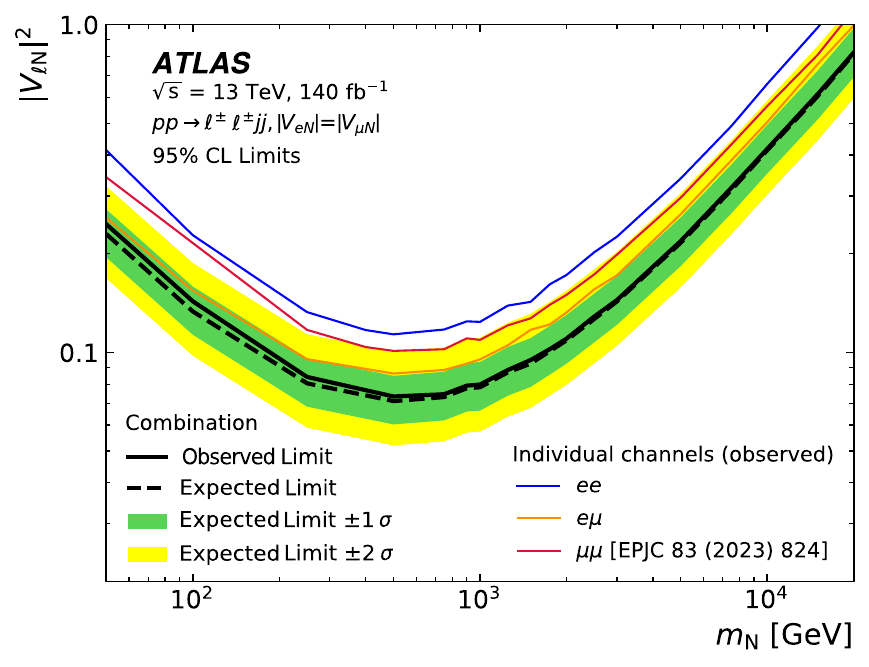}}
\end{minipage}
\hfill
\begin{minipage}{0.34\linewidth}
\centerline{\includegraphics[width=\linewidth]{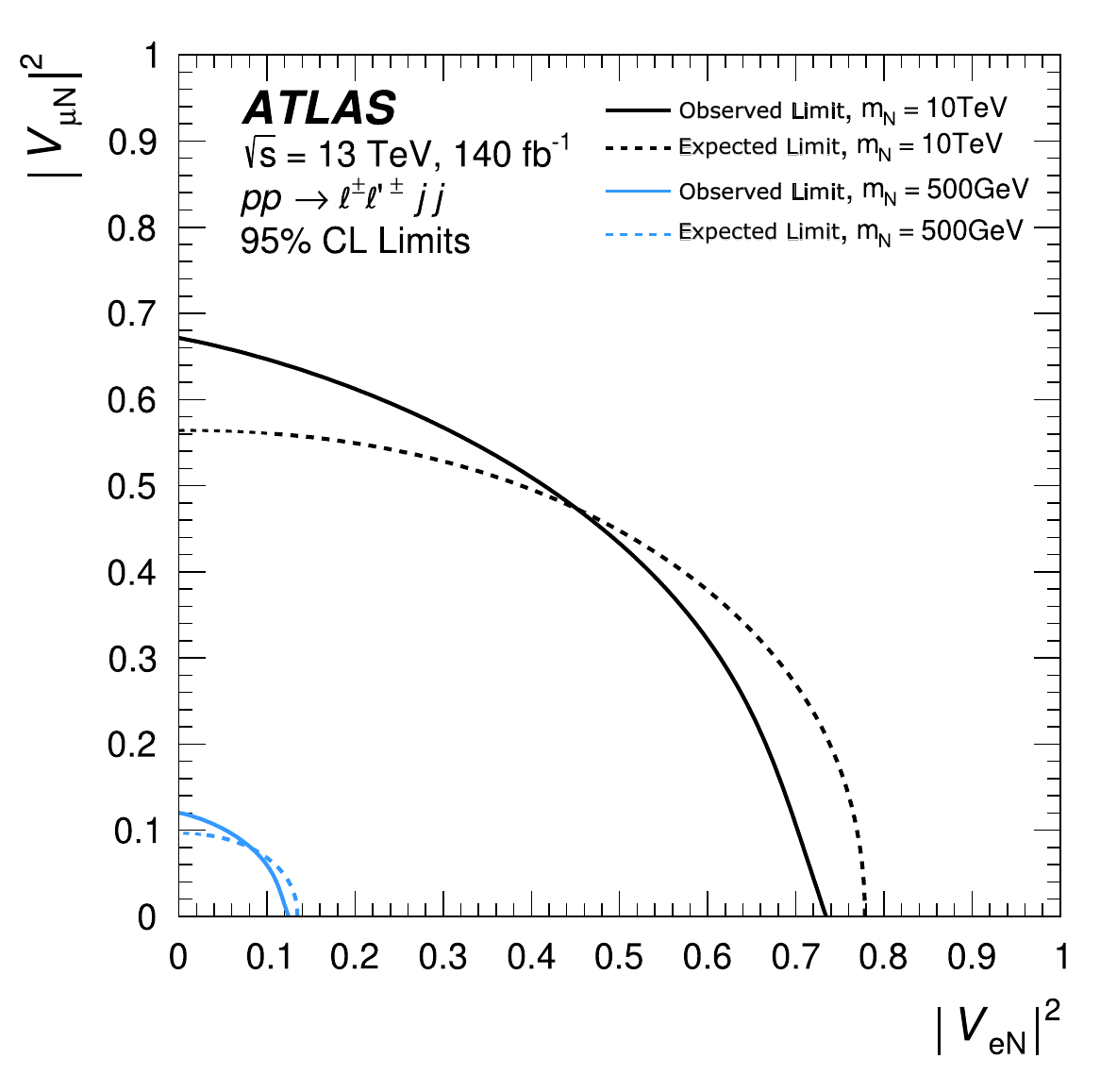}}
\end{minipage}
\caption[]{Production of a Majorana neutrino in $WW$ scattering, with neutrinoless double $\beta$ decay (left). ATLAS limits on neutrino mass $m_N$ vs lepton-neutrino mixing matrix element $|V_{\ell N}|^2$ ($\ell=e,\mu$), at 95\% CL for individual and combined channels (center). Limits on ($|V_{e N}|^2,|V_{\mu N}|^2$) plane (right).\cite{CERN-EP-2024-083}}
\label{fig:EXOT-2023-16}
\end{figure}

\begin{figure}
\begin{minipage}{0.29\linewidth}
\centerline{\includegraphics[width=\linewidth]{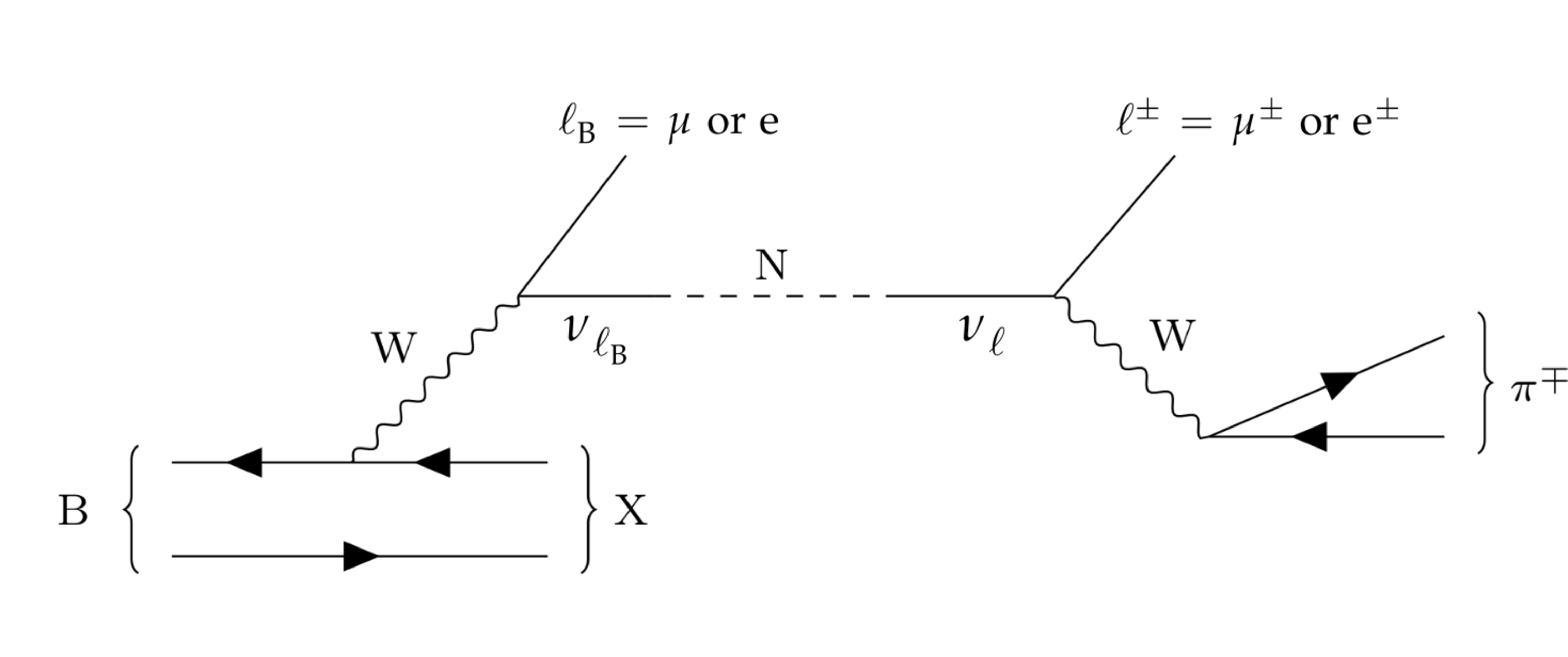}}
\end{minipage}
\hfill
\begin{minipage}{0.34\linewidth}
\centerline{\includegraphics[width=\linewidth]{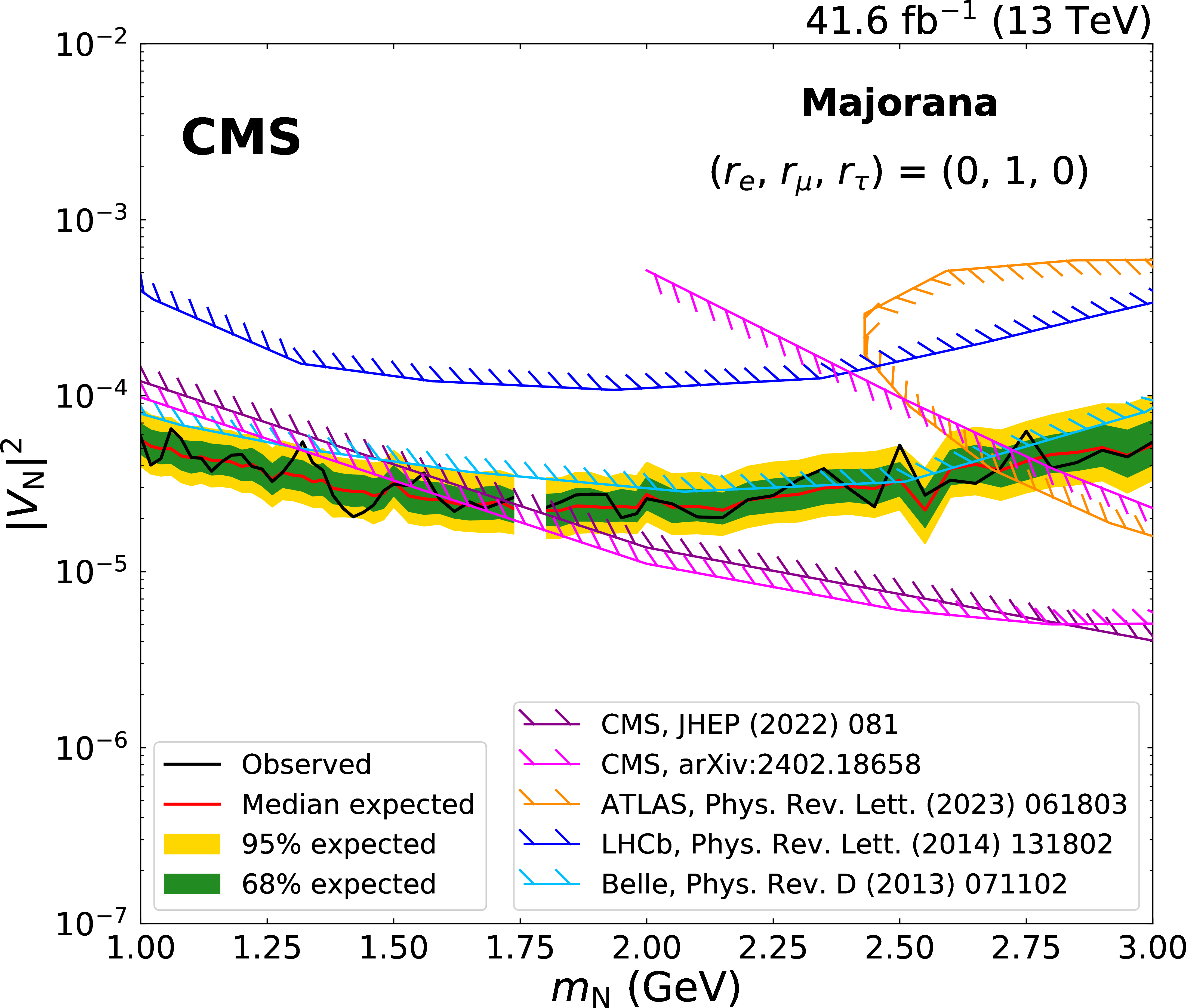}}
\end{minipage}
\hfill
\begin{minipage}{0.34\linewidth}
\centerline{\includegraphics[width=\linewidth]{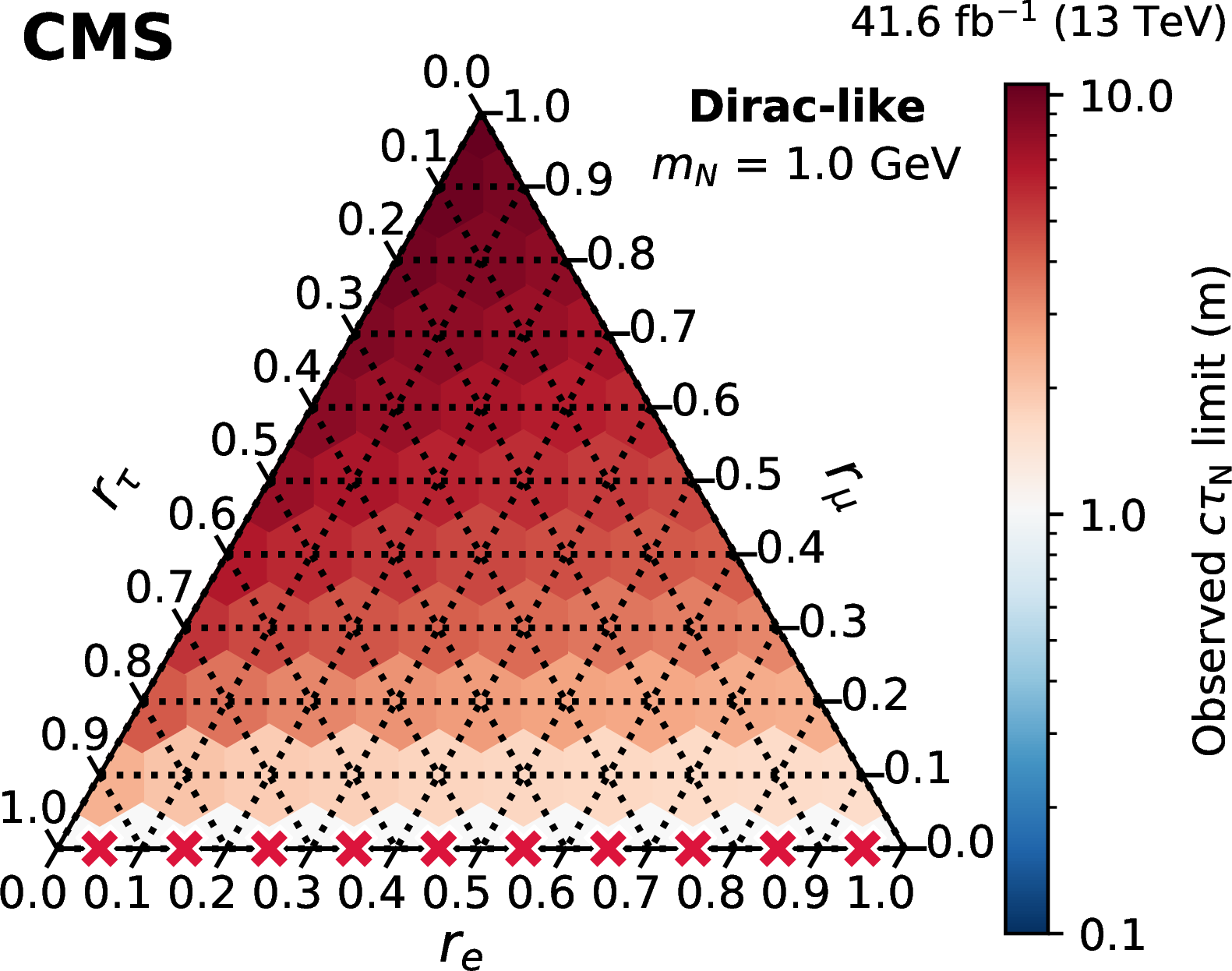}}
\end{minipage}
\caption[]{Production of a heavy neutrino in $B$ meson semi-leptonic decay (left). Fully leptonic channel has been also explored. CMS limits were obtained for several coupling to each family. For exclusive coupling to muons, limits are the most stringent to date (center). Limits on HNL lifetime, as a function of coupling to the three lepton families (right).\cite{CMS-EXO-22-019}}
\label{fig:CMS-EXO-22-019}
\end{figure}

\begin{figure}
\begin{minipage}{0.29\linewidth}
\centerline{\includegraphics[width=\linewidth]{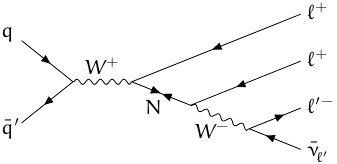}}
\end{minipage}
\hfill
\begin{minipage}{0.34\linewidth}
\centerline{\includegraphics[width=\linewidth]{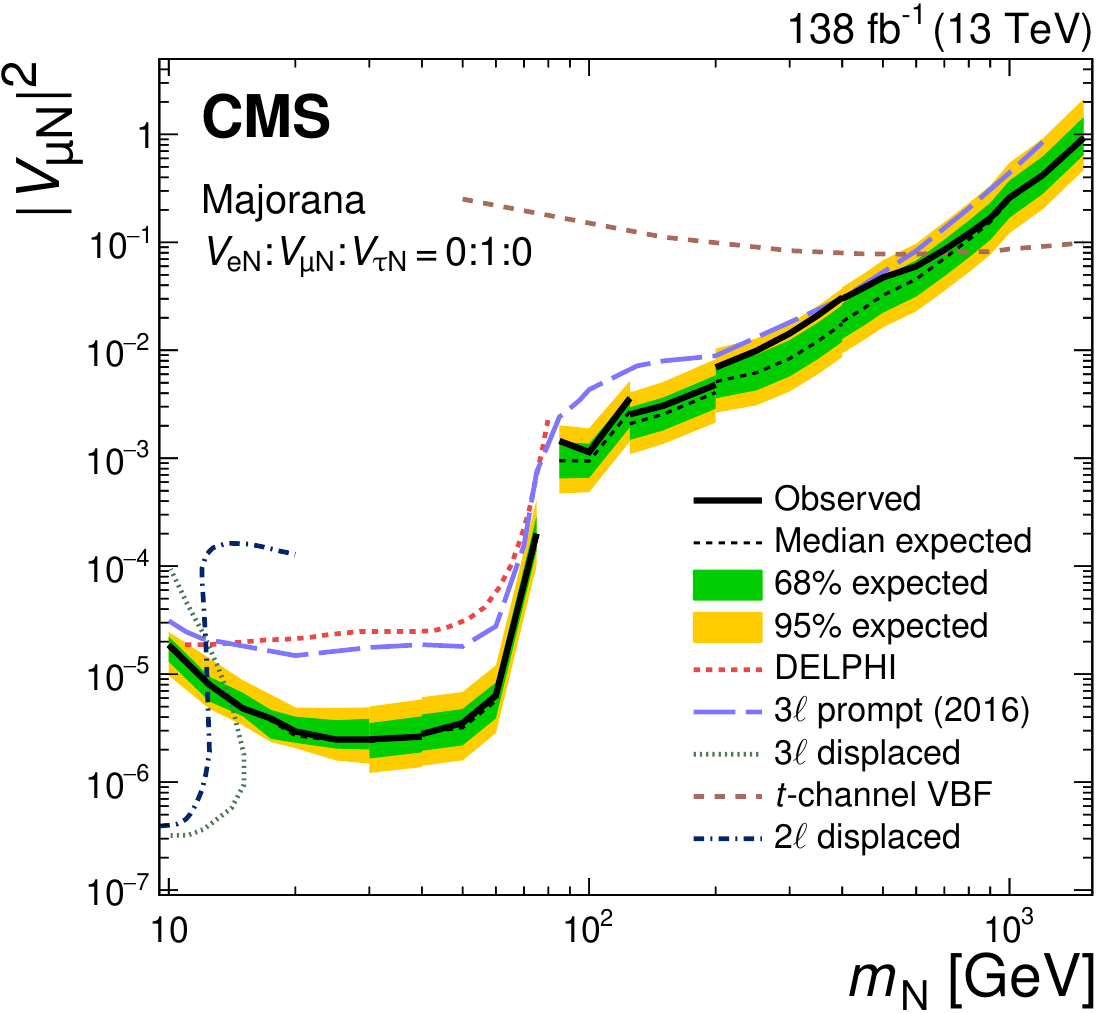}}
\end{minipage}
\hfill
\begin{minipage}{0.34\linewidth}
\centerline{\includegraphics[width=\linewidth]{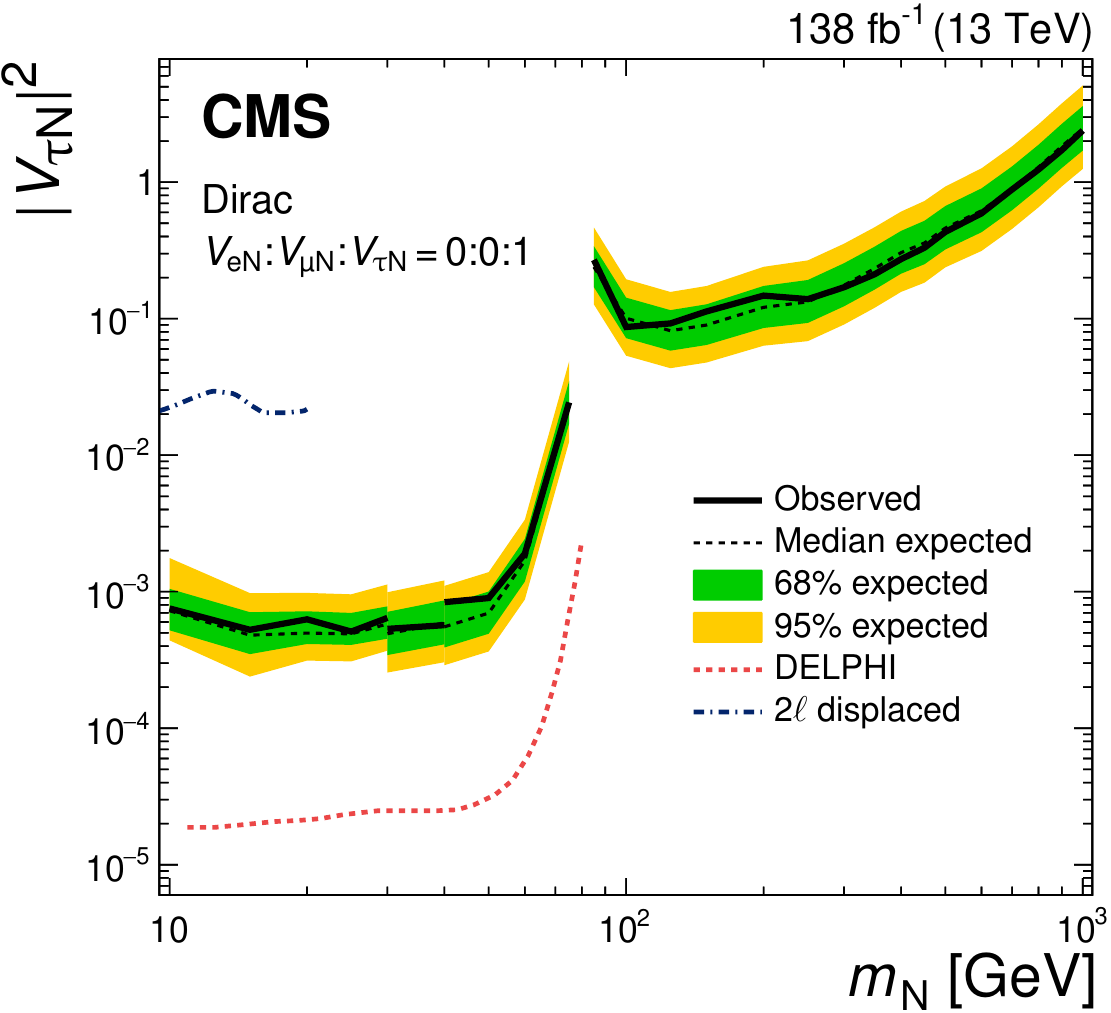}}
\end{minipage}
\caption[]{Production of a three leptons final state ($\ell=e,\mu,\tau$; $\tau\to$hadrons) mediated by an heavy neutrino (left). CMS limits on mixing matrix elements $|V_{\ell N}|^2$ were obtained, e.g. $|V_{\mu N}|^2$ as a function of neutrino mass $m_N$ (center). Limits on mixing matrix element $|V_{\tau N}|^2$ (right), for which the region $m_N>m_W$ is explored for the first time.\cite{CMS-EXO-22-011}}
\label{fig:CMS-EXO-22-011}
\end{figure}

\begin{figure}
\begin{minipage}{0.29\linewidth}
\centerline{\includegraphics[width=\linewidth]{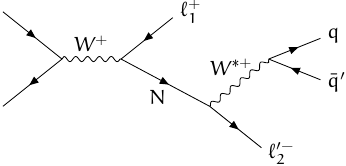}}
\end{minipage}
\hfill
\begin{minipage}{0.34\linewidth}
\centerline{\includegraphics[width=\linewidth]{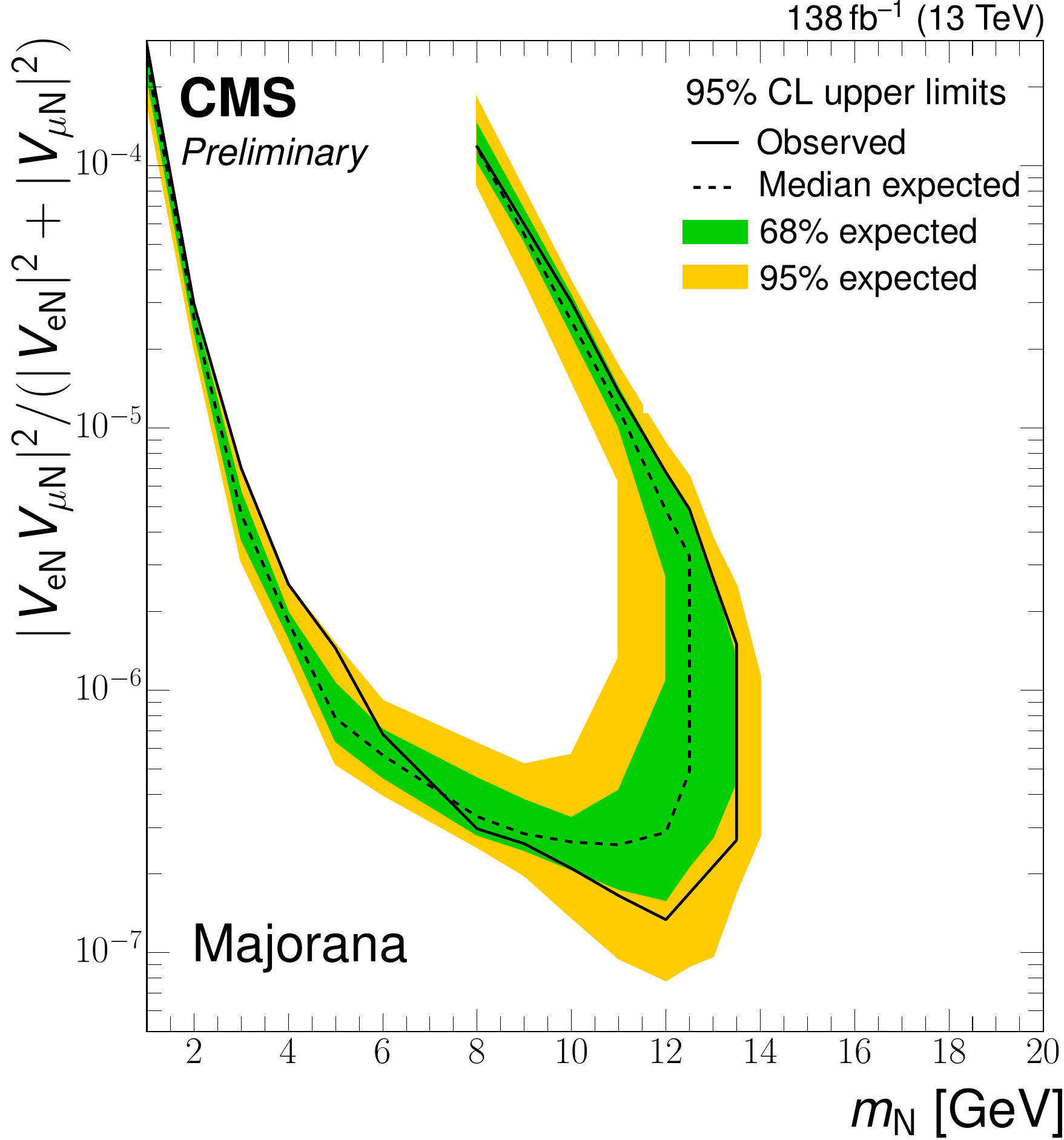}}
\end{minipage}
\hfill
\begin{minipage}{0.34\linewidth}
\centerline{\includegraphics[width=\linewidth]{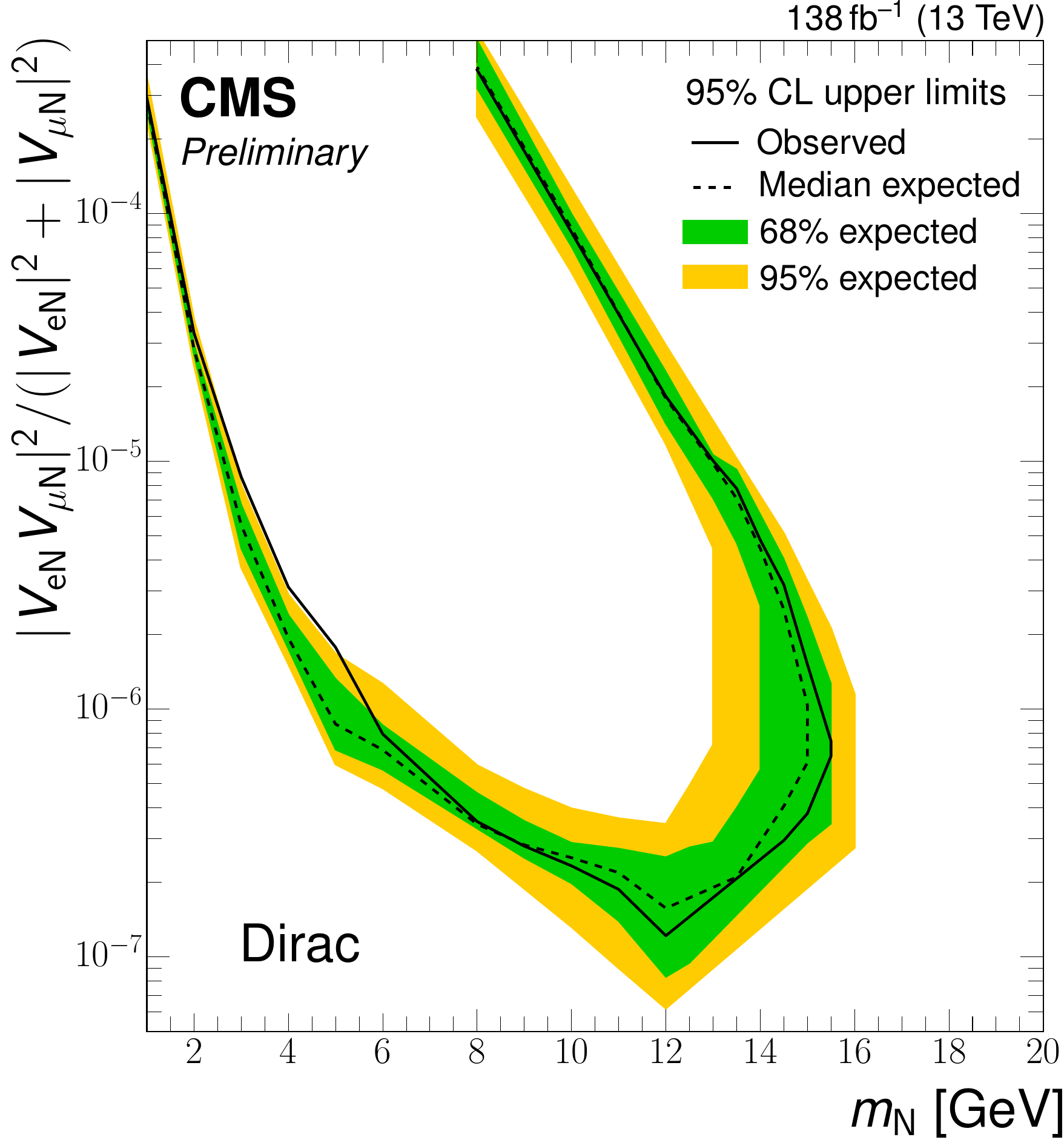}}
\end{minipage}
\caption[]{A heavy neutrino decay, producing a displaced vertex with a lepton and a jet, associated to a prompt lepton (left). CMS exclusion limits on the square mixing parameters, for Majorana (center) and Dirac (right) type. They are the most stringent to date on the range $11 < m_N < 16~\mathrm{GeV}$.\cite{CMS-PAS-EXO-21-011}}
\label{fig:CMS-EXO-21-011}
\end{figure}

\begin{figure}
\begin{minipage}{0.48\linewidth}
\centerline{\includegraphics[width=\linewidth]{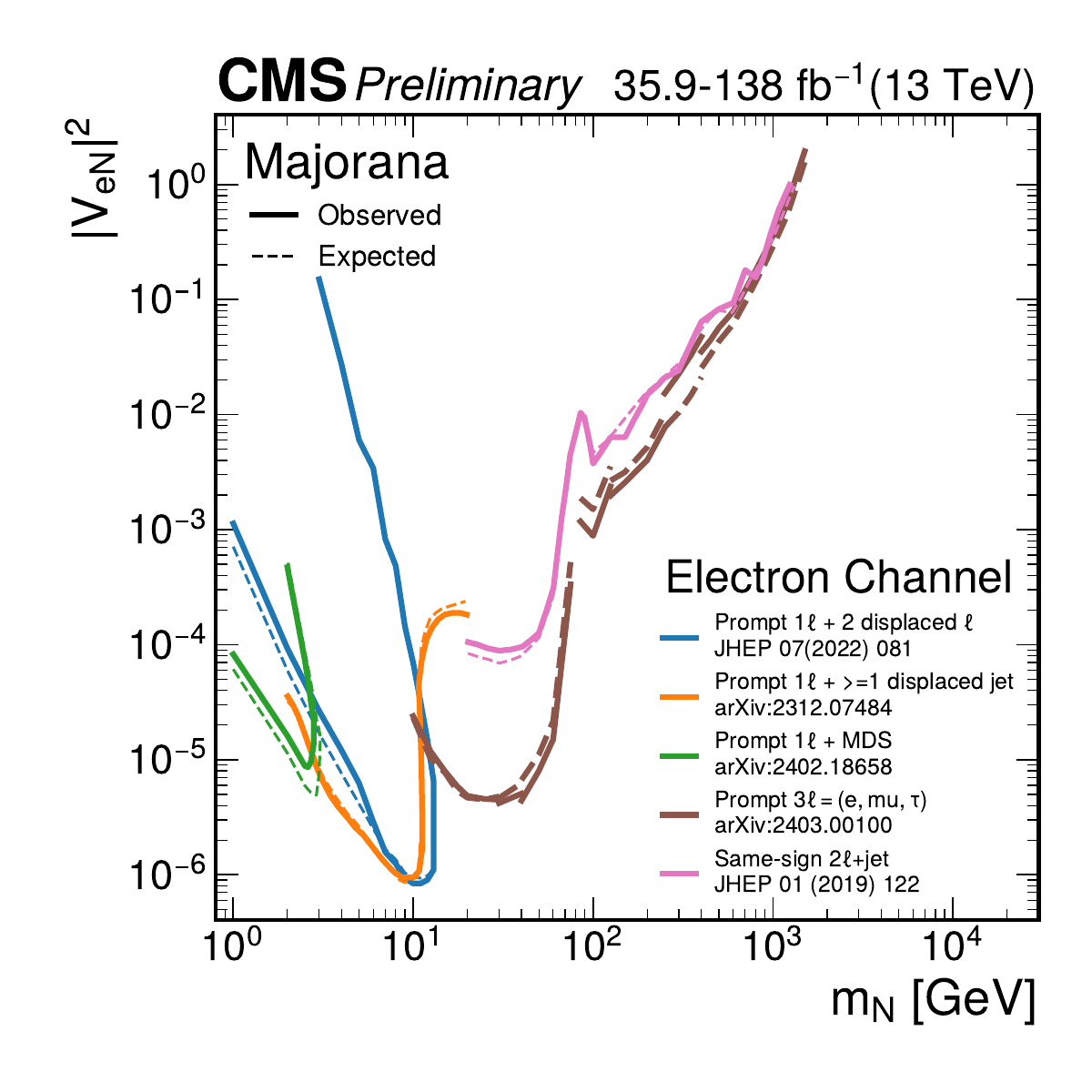}}
\end{minipage}
\hfill
\begin{minipage}{0.48\linewidth}
\centerline{\includegraphics[width=\linewidth]{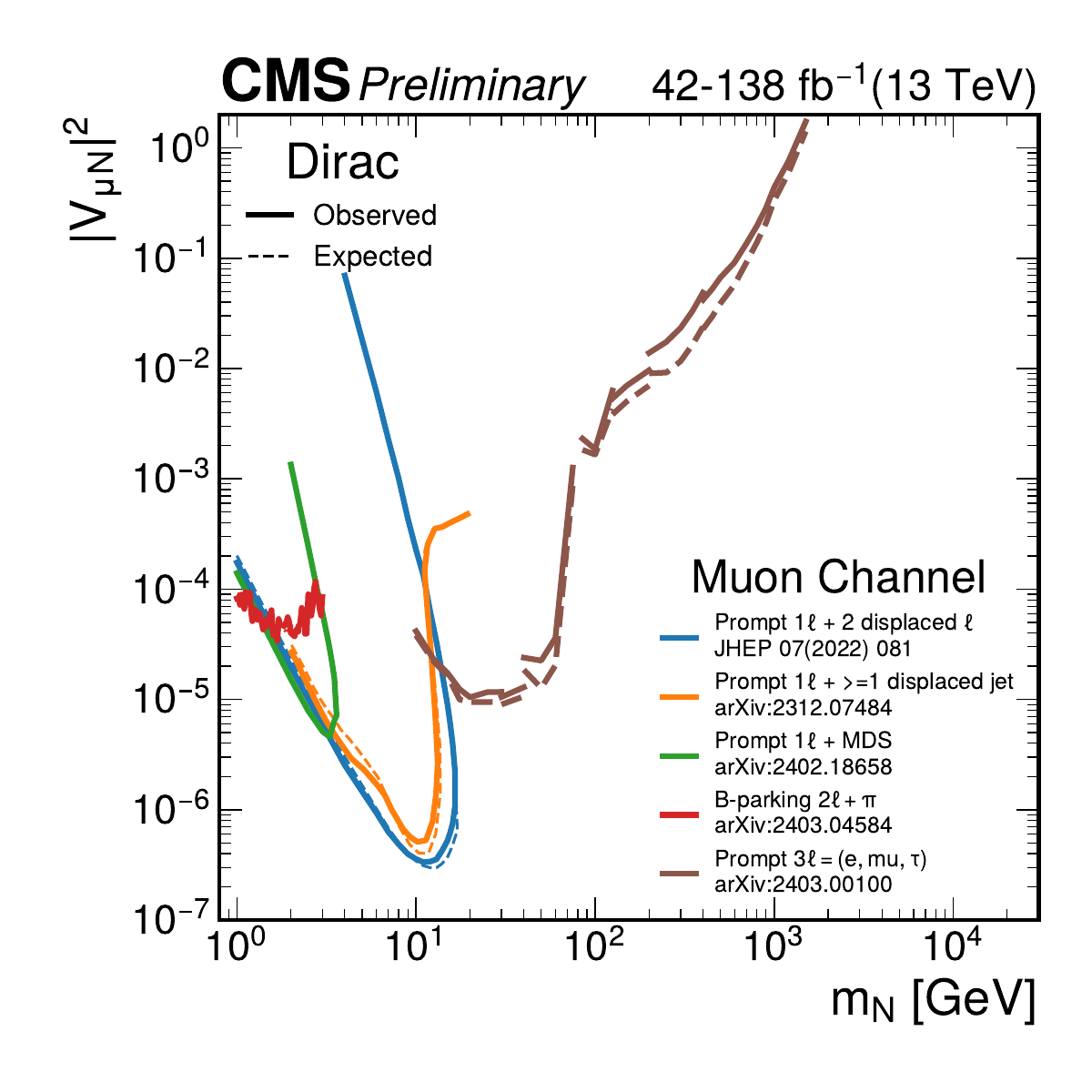}}
\end{minipage}
\caption[]{CMS summary of the HNL exclusion limits for a Majorana neutrino in the $ee$ lepton channel (left), and for a Dirac neutrino in the $\mu\mu$ channel (right).\cite{CMS-HNL-summary}}
\label{fig:CMS-HNL-summary}
\end{figure}

Many BSM theories predict the existence of heavy neutrinos $N$, mixing with SM lepton families ($\ell$) via matrix elements $|V_{\ell N}|$. Their lifetime is releated to their mass as $\tau_N\sim |V_{N}|^{-2}m_N^{-5}$. They could be of Dirac type, leading to lepton number conserving final states, or Majorana type, that allows also lepton number violation. The results for the explored channels are presented in Figure~\ref{fig:EXOT-2023-16} for ATLAS, and in Figures~\ref{fig:CMS-EXO-22-019}-\ref{fig:CMS-EXO-21-011} for CMS. A summary including other CMS publications, on HNL exclusion limits, is shown in Figure~\ref{fig:CMS-HNL-summary}.

\section{Long-lived particles}\label{Section:LLP}
\begin{figure}
\begin{minipage}{0.22\linewidth}
\centerline{\includegraphics[width=\linewidth]{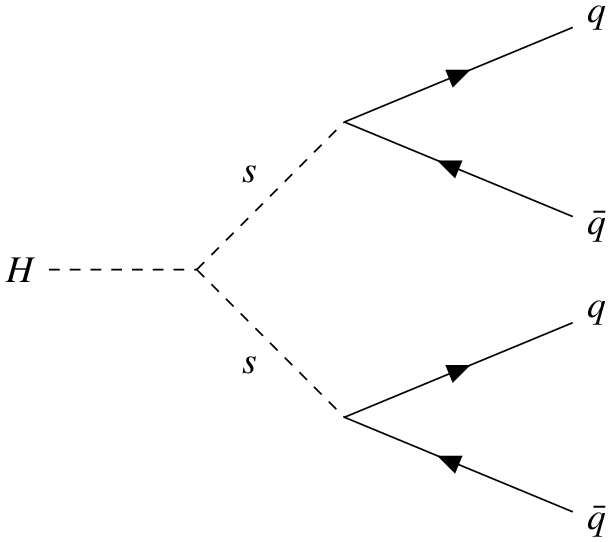}}
\end{minipage}
\hfill
\begin{minipage}{0.22\linewidth}
\centerline{\includegraphics[width=\linewidth]{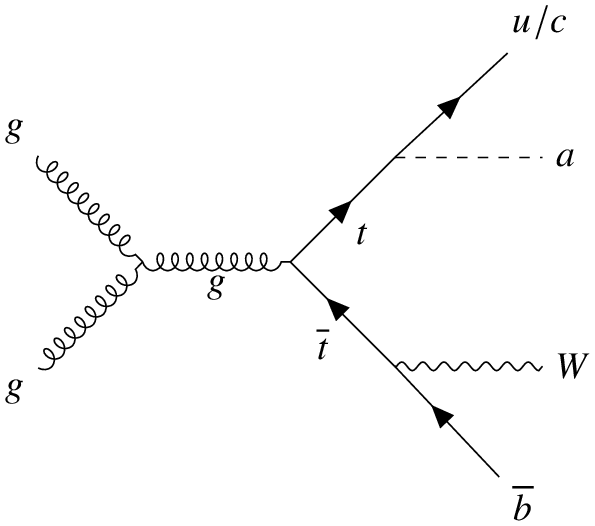}}
\end{minipage}
\hfill
\begin{minipage}{0.53\linewidth}
\centerline{\includegraphics[width=\linewidth]{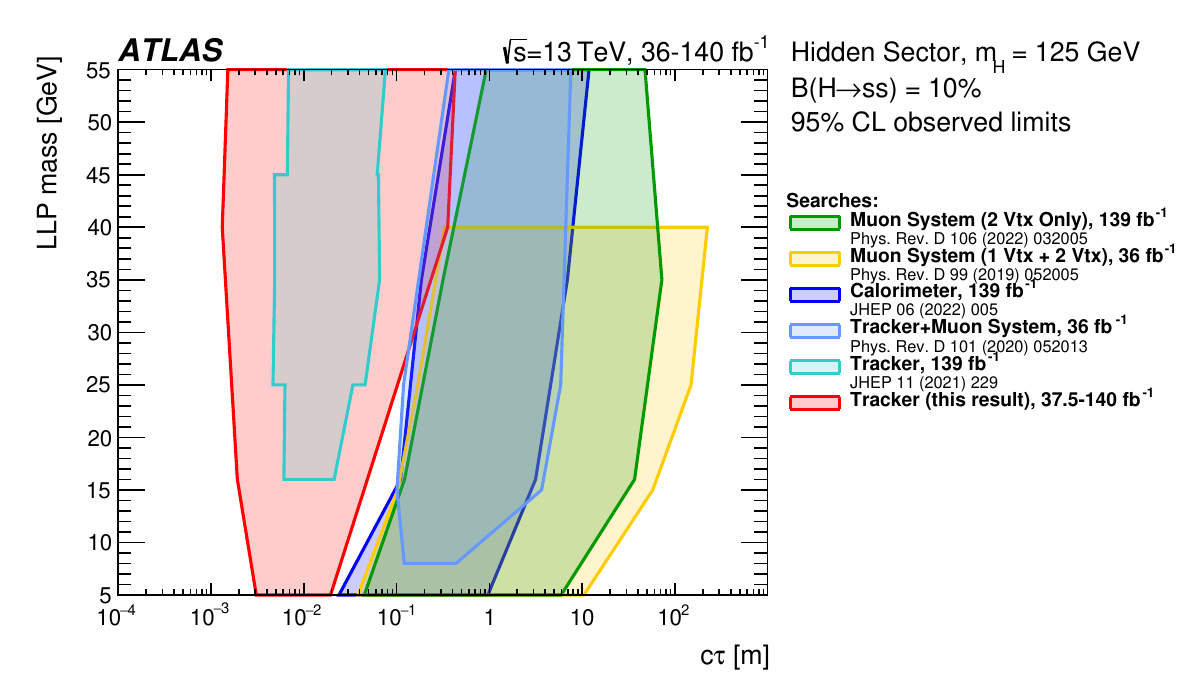}}
\end{minipage}
\caption[]{Two LLP benchmark signatures shown: a pair of long-lived neutral pseudoscalar bosons $s$, produced through an ``Higgs portal'', going into displaced jets (left); a pseudo Nambu-Goldstone boson axion-like particle $a$, associated to a vector boson, probed for the first time (center). Exclusion limits in the LLP mass vs lifetime plane from ATLAS searches using a ${\cal B}(H\to ss)=10\%$. Limits for final states with $4b$ were improved ${\cal O}(10)$ times, while those with $4c$ were probed for the first time (right).\cite{EXOT-2021-32}}
\label{fig:EXOT-2021-32}
\end{figure}

\begin{figure}
\begin{minipage}{0.37\linewidth}
\centerline{\includegraphics[width=\linewidth]{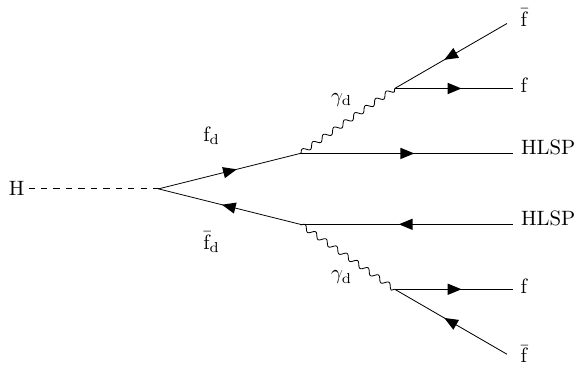}}
\end{minipage}
\hfill
\begin{minipage}{0.6\linewidth}
\centerline{\includegraphics[width=\linewidth]{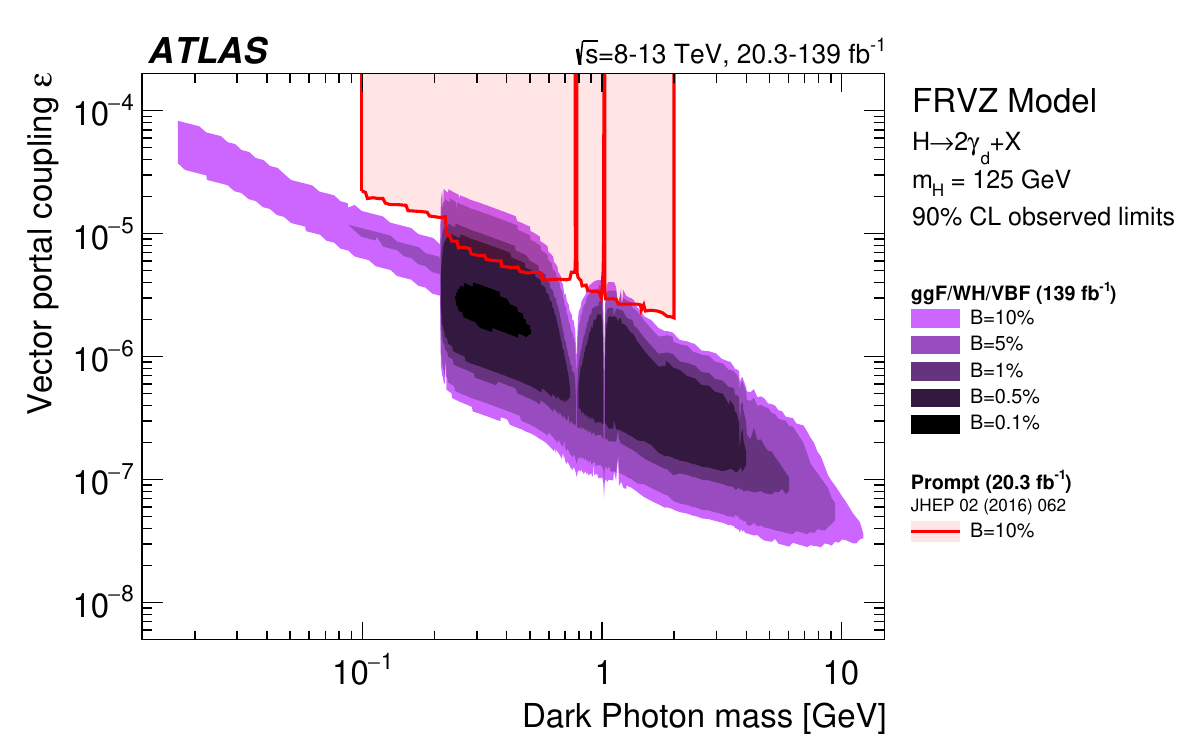}}
\end{minipage}
\caption[]{Production of a dark photon pair in a Higgs decay. The final state consists of SM fermions and hidden lightest stable particles - HLSP (left). ATLAS limits on the kinetic mixing parameter as a function of the dark photon mass (right).\cite{EXOT-2022-15}}
\label{fig:EXOT-2022-15}
\end{figure}

\begin{figure}
\begin{minipage}{0.32\linewidth}
\centerline{\includegraphics[width=\linewidth]{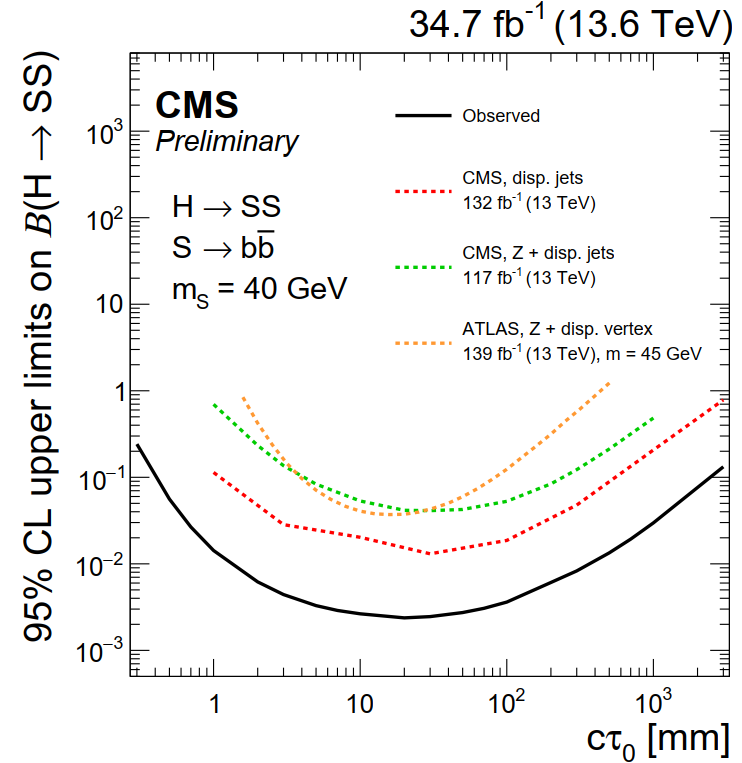}}
\end{minipage}
\hfill
\begin{minipage}{0.32\linewidth}
\centerline{\includegraphics[width=\linewidth]{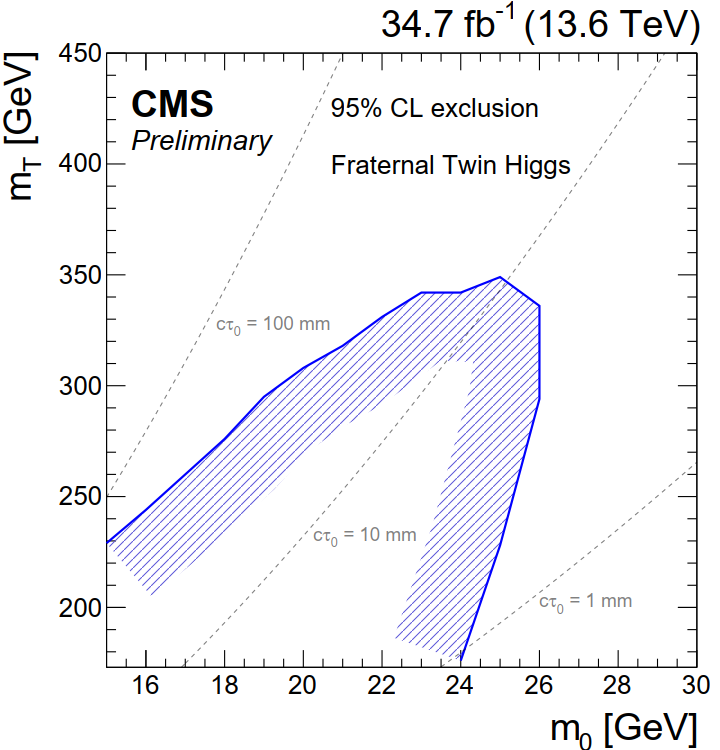}}
\end{minipage}
\begin{minipage}{0.34\linewidth}
\centerline{\includegraphics[width=\linewidth]{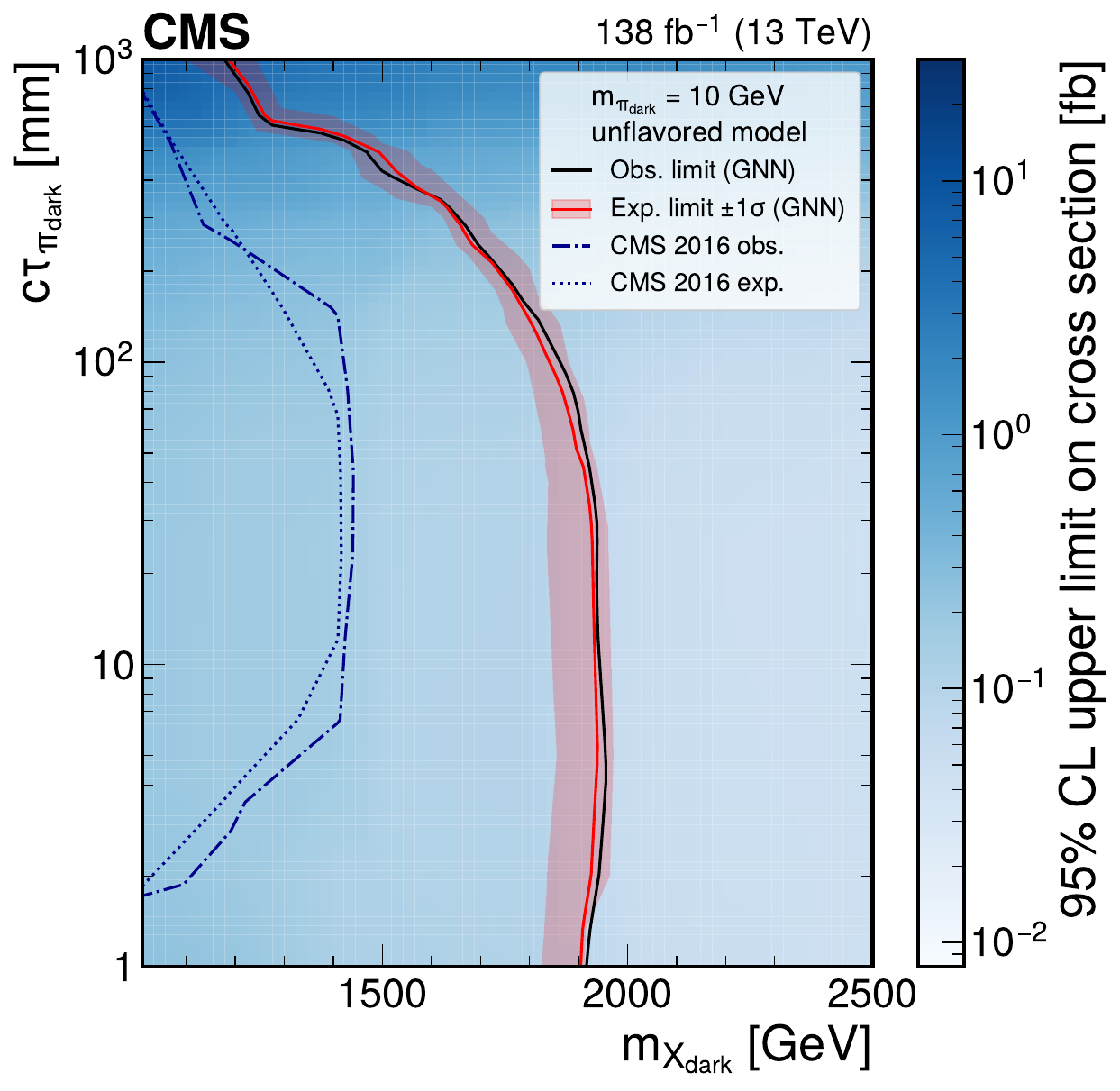}}
\end{minipage}
\caption[]{Using recent data collected at $13.6~\mathrm{TeV}$, the same ``Higgs portal'' signature of Figure~\ref{fig:EXOT-2021-32}, but considering also $\tau$ leptons. Hadronic $\tau$ lepton decays were probed for the first time in the range $c\tau < 1$~m. CMS set limits on the ${\cal B}(H\to ss)$ vs LLP lifetime (left). Mass limits were set in a Fraternal Twin Higgs model in which $s$ is a hidden glueball of mass $m_0$, and $m_T$ is the mass of the top-partner in the hidden sector (center).\cite{CMS-PAS-EXO-23-013} Another hypothetical scalar mediator $X_{dark}$, produced in pairs via $gg$ fusion or $q\bar{q}$ annihilation, could decay into SM and dark quarks, producing jets with multiple displaced vertices. Based on this signature, limits were set on $X_{dark}$ mass as a function of dark meson lifetime, here shown for $m_{\pi_{dark}}=10~\mathrm{GeV}$ (right).\cite{CMS:2024gxp}}
\label{fig:CMS-EXO-23-013}
\end{figure}

\begin{figure}
\begin{minipage}{0.32\linewidth}
\centerline{\includegraphics[width=\linewidth]{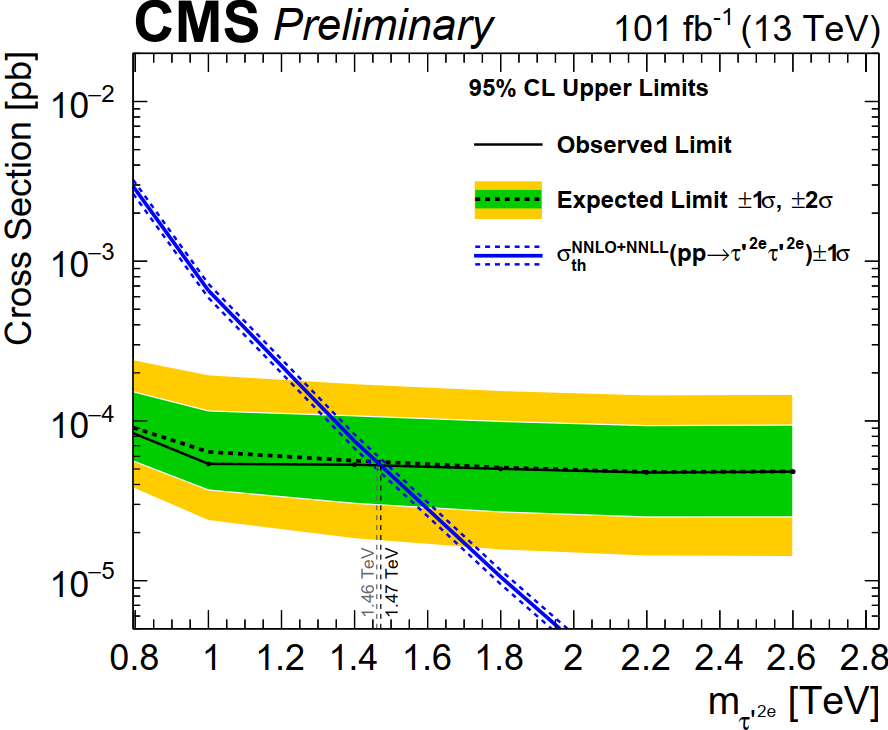}}
\end{minipage}
\hfill
\begin{minipage}{0.32\linewidth}
\centerline{\includegraphics[width=\linewidth]{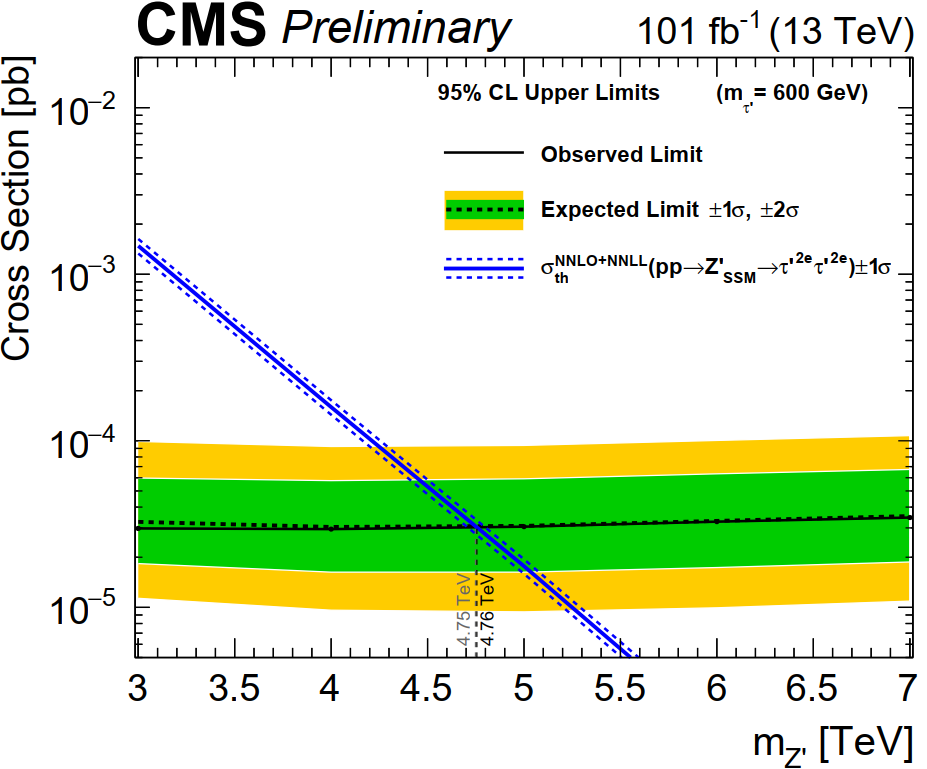}}
\end{minipage}
\hfill
\begin{minipage}{0.34\linewidth}
\centerline{\includegraphics[width=\linewidth]{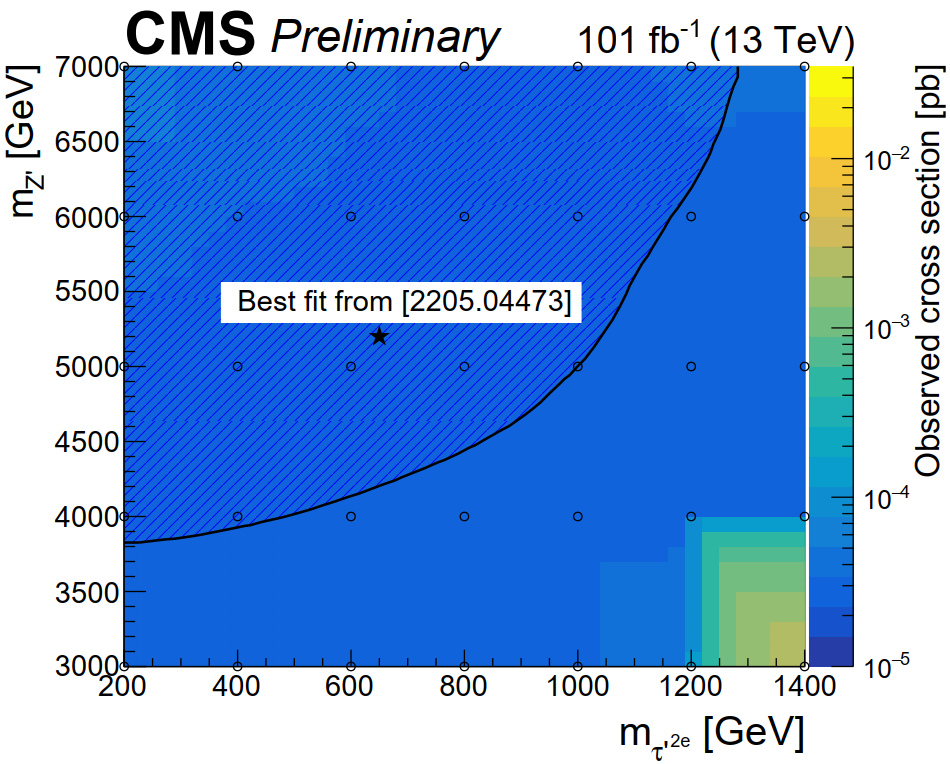}}
\end{minipage}
\caption[]{CMS exclusion limits on the mass of a $\tau^\prime$ lepton with charge $|Q|=2e$ obtained looking for large ionization energy losses in the silicon inner detector (left). Exclusion limits on a heavy $Z^\prime$ boson decaying in two $\tau^\prime$ assuming $m_{\tau^\prime}=600~\mathrm{GeV}$, $|Q|=2e$ (center). Exclusion limits in the $(m_{Z^\prime}, m_{\tau^\prime})$ plane (right).\cite{CMS-PAS-EXO-18-002}}
\label{fig:CMS-EXO-18-002}
\end{figure}

Hypothetical BSM long-lived particles would produce peculiar detector signatures, involving e.g. displaced vertices, different jet shapes, etc., which all needed to develop specific search strategies. Results from the searches for these are shown in Figures~\ref{fig:EXOT-2021-32}-\ref{fig:EXOT-2022-15} in case of ATLAS and in Figures~\ref{fig:CMS-EXO-23-013}-\ref{fig:CMS-EXO-18-002} in case of CMS.

\section{Vector-like quarks}\label{Section:VLQ}
\begin{figure}
\begin{minipage}{0.23\linewidth}
\centerline{\includegraphics[width=\linewidth]{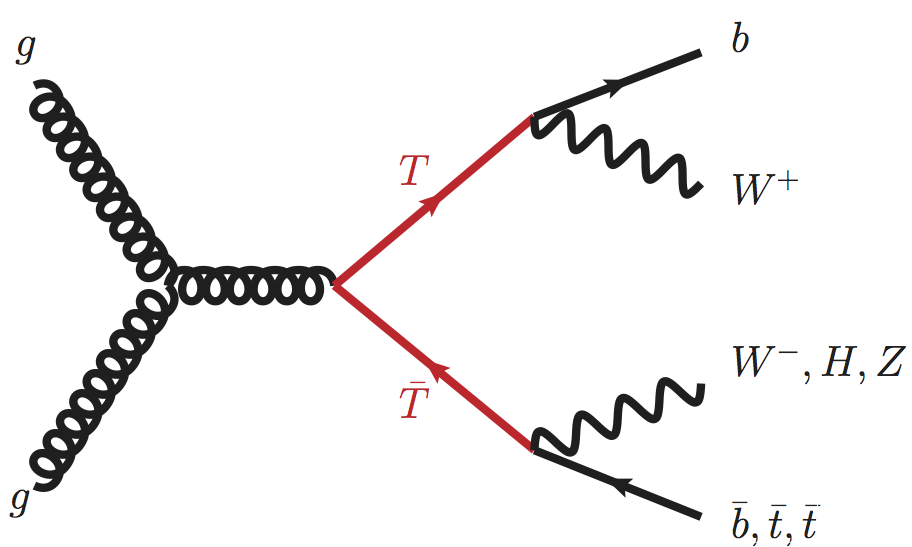}}
\end{minipage}
\hfill
\begin{minipage}{0.38\linewidth}
\centerline{\includegraphics[width=\linewidth]{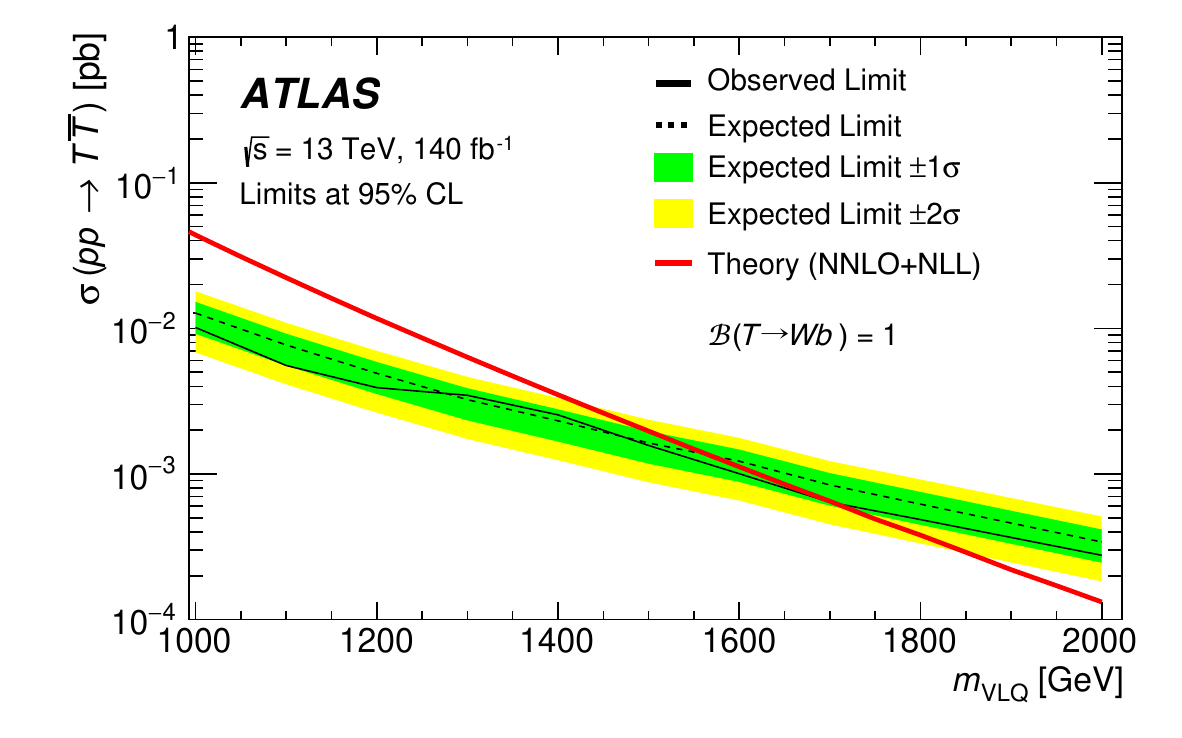}}
\end{minipage}
\hfill
\begin{minipage}{0.35\linewidth}
\centerline{\includegraphics[width=\linewidth]{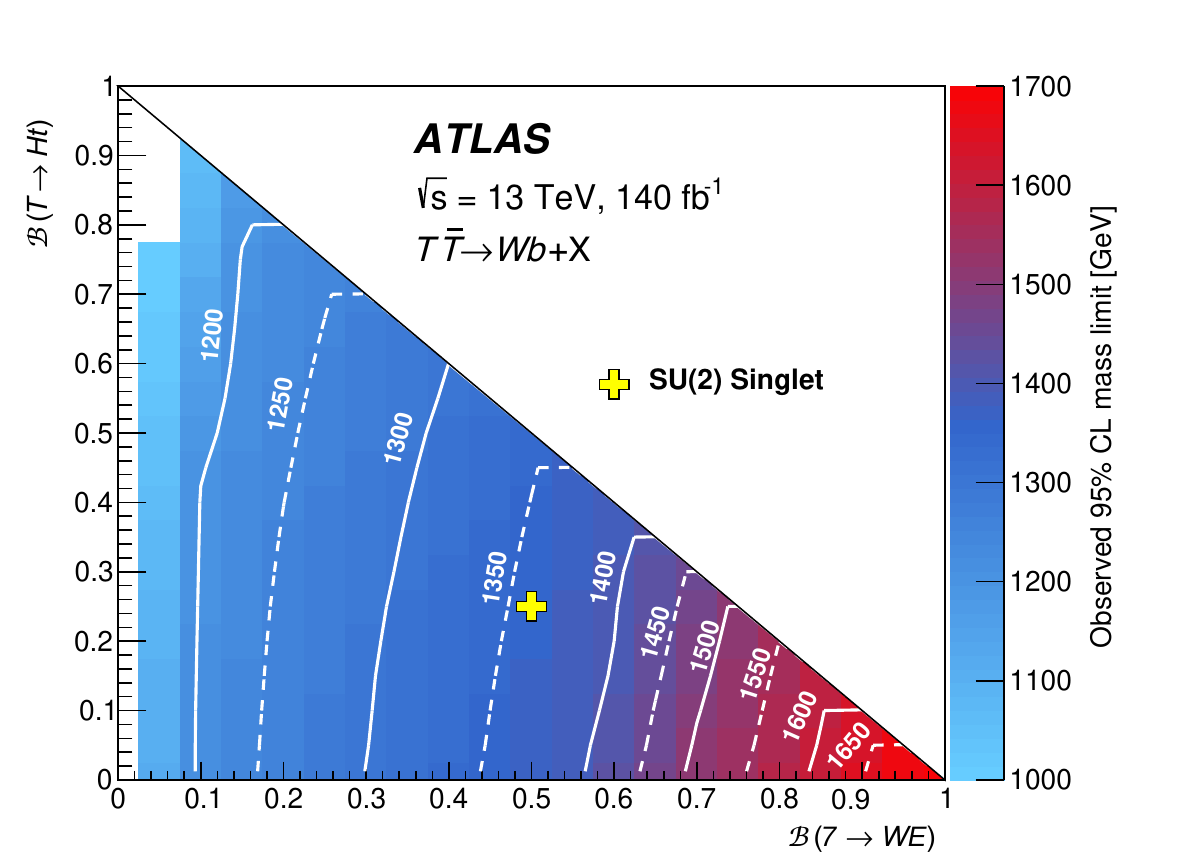}}
\end{minipage}
\hfill
\caption[]{Optimized for ${\cal B}(T\to Wb)=1$ (left), but sensitive to all VLQs decays, ATLAS set limits on the $m_{VLQ}$ mass (center). Observed limits in the plane ${\cal B}(T\to Wb)$ vs ${\cal B}(T\to Ht))$ (right).\cite{EXOT-2019-06}}
\label{fig:EXOT-2019-06}
\end{figure}

\begin{figure}
\begin{minipage}{0.24\linewidth}
\centerline{\includegraphics[width=\linewidth]{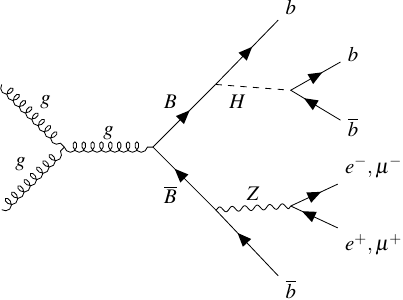}}
\end{minipage}
\hfill
\begin{minipage}{0.33\linewidth}
\centerline{\includegraphics[width=\linewidth]{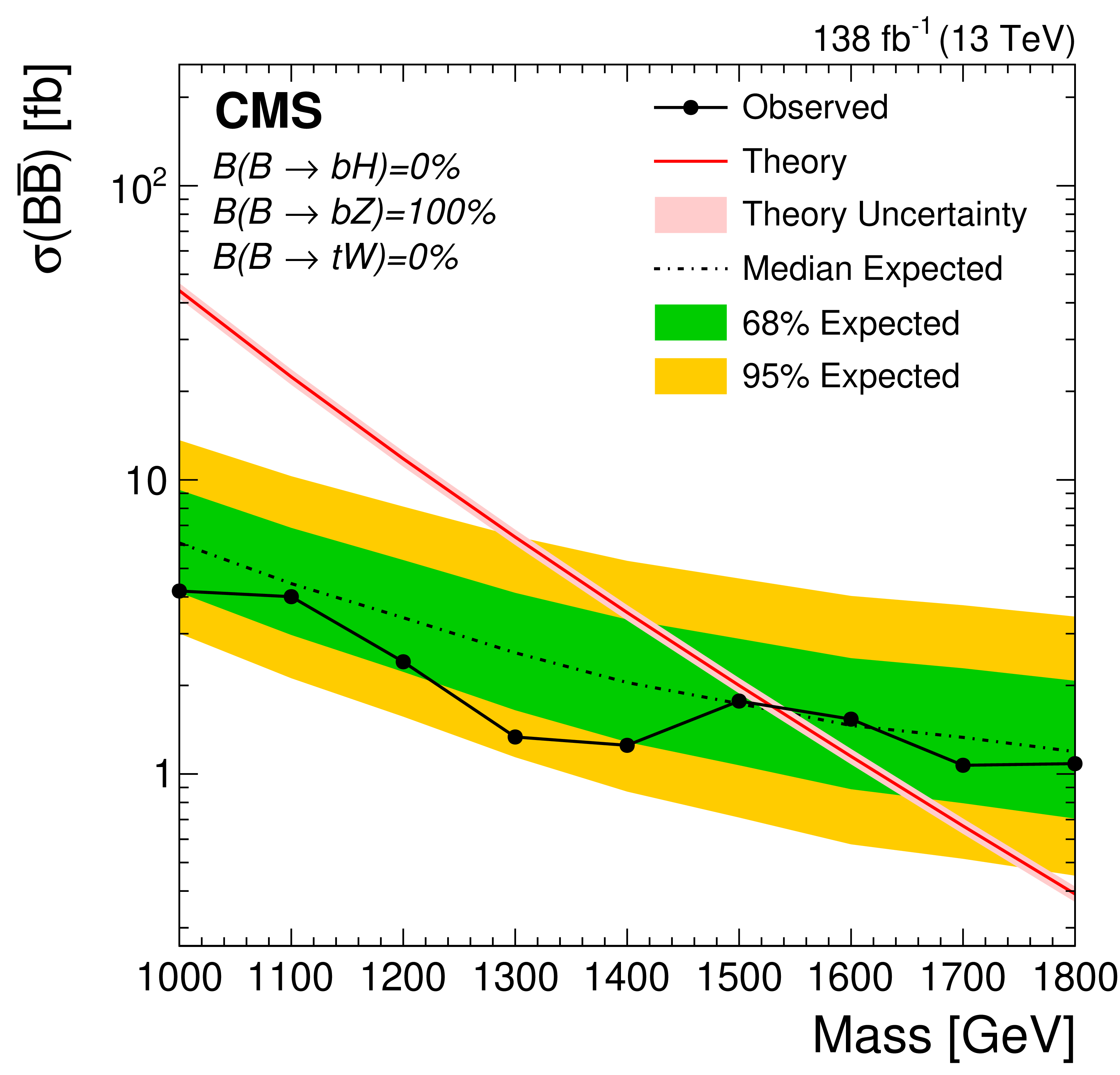}}
\end{minipage}
\hfill
\begin{minipage}{0.41\linewidth}
\centerline{\includegraphics[width=\linewidth]{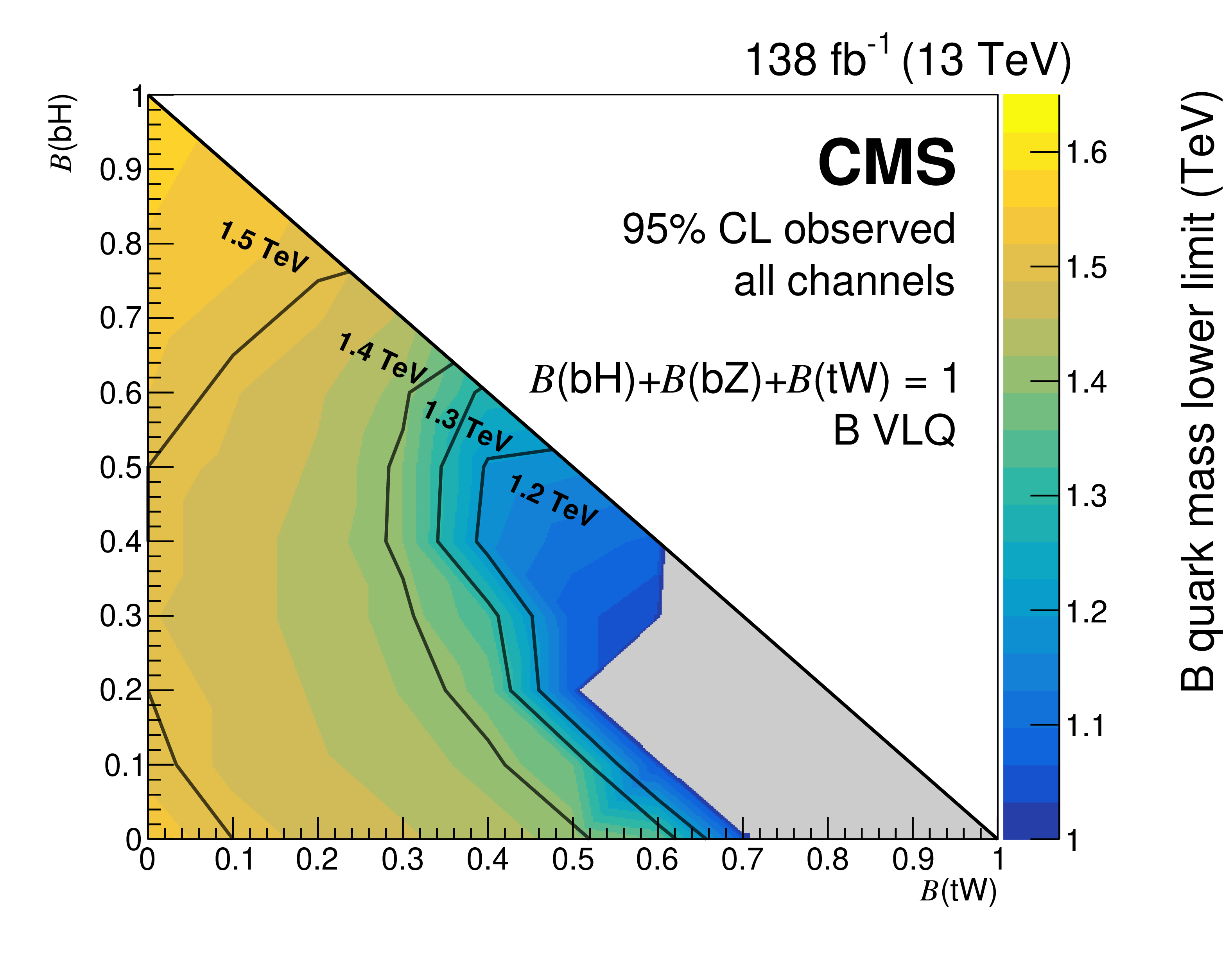}}
\end{minipage}
\caption[]{Using the number of reconstructed jets in fully hadronic final states, including leptonic $Z$ decays (left), CMS set limits on different branching fraction hypothesis for the vector-like $B$. Case for ${\cal B}(B\to bZ)=1$ (center), and observed exclusion limit in ${\cal B}(B\to bH)$ vs ${\cal B}(B\to tW)$  (right.)\cite{CMS-B2G-20-014}}
\label{fig:CMS-B2G-20-014}
\end{figure}

VLQs are introduced by models like Little/Composite Higgs, large extra-dimensions, string theory, to allow a mass term directly added in the Lagrangian $\cal L$, being independent of Yukawa couplings to $H$. They would mix mainly to third generation quarks, to cancel $H$ mass divergences. VLQs can be in singlet, doublet or triplet. The vector-like $T$ and $B$ have electric charge $2/3$ and $-1/3$, i.e. the same charge of the corresponding SM partners top and bottom quarks. Vector-like quarks with exotic charges could also exist, such as the $X$ and $Y$ quarks, with $+5/3$ and $-4/3$, respectively. They can decay as: $T\to tH$, $tZ$, $bW$; $Y\to bW$; $B\to bH$, $bZ$, $tW$; $X\to tW$. Last results on $B$ and $T$ searches from ATLAS and CMS are shown in Figure~\ref{fig:EXOT-2019-06} and in Figure~\ref{fig:CMS-B2G-20-014}, respectively.

\section{Conclusions}
ATLAS and CMS searched for several new particles, including heavy
neutral leptons, long-lived particles, and vector-like quarks. No
signal of physics BSM has been observed, extending further the
exclusion limits for the analyzed scenarios.
Regarding HNL: in $WW$ boson scattering, Majorana neutrino masses  $|m_{ee}|<24.5~\mathrm{GeV}$, $|m_{e\mu}|<12.5~\mathrm{GeV}$ are excluded for $ee, e\mu$ final states;~\cite{CERN-EP-2024-083} in $B$ meson decays to light neutrino, the region $(|V_{eN}|^2+|V_{\mu N}|^2+|V_{\tau N}|^2)<2.0\cdot 10^{-5}$, $c\tau_N<10.5$~m is excluded;~\cite{CMS-EXO-22-019} in the three leptons final states, the exclusion range in the plane $(m_N,|V_{eN}|^2)$ is improved, and the region $|V_{\tau N}|^2$ ($m_N>m_W$) is explored for the first time;~\cite{CMS-EXO-22-011} for a long-lived $N$ in the $\ell +$jets channel, best limit are set to mixing parameters in $11 < m_N < 16~\mathrm{GeV}$ range;~\cite{CMS-PAS-EXO-21-011} overall summary is updated accordingly.~\cite{CMS-HNL-summary}
Regarding LLP: using displaced inner detector vertices, limits to $a,s$ bosons production were set for different masses and lifetimes, not explored before;~\cite{EXOT-2021-32}
long-lived dark photons in Higgs boson decays are excluded for $\cal B$(H$\to 2\gamma_d + $X)$>10\%$ with $173 < c\tau_{\gamma_d} <1296$~mm;~\cite{EXOT-2022-15}
using reconstruction of displaced jets, limits for Higgs decaying to neutral scalars are set for $15<m_{LLP}<55~\mathrm{GeV}$, $c\tau<1$;~\cite{CMS-PAS-EXO-23-013}
exclusion limits are set in the $(m_{X_{dark}}$, $c\tau_{\pi_{dark}})$ plane for long-lived dark mesons with $m_{\pi_{dark}}\sim 10~\mathrm{GeV}$;~\cite{CMS:2024gxp}
looking for large silicon $dE/dx$ deposit, $m_{Z^\prime_{SSM}}<4.76~\mathrm{GeV}$ are excluded for $m_{\tau^{\prime(2e)}}=600~\mathrm{GeV}$.~\cite{CMS-PAS-EXO-18-002}
Regarding VLQ: vector-like $T\bar{T}$ and $B\bar{B}$ are excluded in the ${\cal B}(T\to Wb), {\cal B}(T\to Ht)$ and ${\cal B}(B\to tW), {\cal B}(B\to bH)$ planes, respectively.~\cite{EXOT-2019-06}$^,$\cite{CMS-B2G-20-014}
New data is being collected at the LHC and, meanwhile, new complex
analysis techniques are being developed by the ATLAS and CMS
Collaborations, in order to improve sensitivity to a wide set of new
physics signals.

\section*{References}

\bibliography{grancagnolo}

\begin{thebibliography}{10}

\bibitem{PERF-2007-01}
{ATLAS Collaboration}.
\newblock {The ATLAS Experiment at the CERN Large Hadron Collider}.
\newblock {\em JINST}, 3:S08003, 2008.

\bibitem{CMS-CMS-00-001}
{CMS Collaboration}.
\newblock {The CMS Experiment at the CERN LHC}.
\newblock {\em JINST}, 3:S08004, 2008.

\bibitem{CERN-EP-2024-083}
{ATLAS Collaboration}.
\newblock {Search for heavy Majorana neutrinos in $e^{\pm} e^{\pm}$ and
  $e^{\pm} \mu^{\pm}$ final states via $WW$ scattering in $pp$ collisions at
  $\sqrt{s}=13$ TeV with the ATLAS detector}.
\newblock {\em CERN-EP-2024-083, submitted to {\em Phys. Lett. B}}, 2024.
\newblock \url{https://arxiv.org/abs/2403.15016}.

\bibitem{CMS-EXO-22-019}
{CMS Collaboration}.
\newblock {Search for long-lived heavy neutrinos in the decays of \(B\) mesons
  produced in proton--proton collisions at \(\sqrt{s} = 13\,\text{TeV}\)}.
\newblock {\em CERN-EP-2024-053, submitted to JHEP}, 2024.
\newblock \url{http://arxiv.org/abs/arXiv:2403.04584}.

\bibitem{CMS-EXO-22-011}
{CMS Collaboration}.
\newblock {Search for heavy neutral leptons in final states with electrons,
  muons, and hadronically decaying tau leptons in proton--proton collisions at
  \(\sqrt{s}\) =\(13\,\text{TeV}\)}.
\newblock {\em CERN-EP-2024-032, submitted to JHEP}, 2024.
\newblock \url{http://arxiv.org/abs/arXiv:2403.00100}.

\bibitem{CMS-PAS-EXO-21-011}
{CMS Collaboration}.
\newblock {Search for long-lived heavy neutral leptons in proton-proton
  collision events with a lepton and a jet from a secondary vertex at
  $\sqrt{s}=13\,\mathrm{TeV}$}.
\newblock {\em CMS-PAS-EXO-21-011}, 2024.
\newblock \url{https://cds.cern.ch/record/2892670}.

\bibitem{CMS-HNL-summary}
{CMS Collaboration}.
\newblock {Summary on the mixing parameter squared as a function of the HNL
  mass}.
\newblock
  \url{https://twiki.cern.ch/twiki/bin/view/CMSPublic/SummaryPlotsEXO13TeV}.

\bibitem{EXOT-2021-32}
{ATLAS Collaboration}.
\newblock {Search for light long-lived particles in \(pp\) collisions at
  \(\sqrt{s}=13\) TeV using displaced vertices in the ATLAS inner detector}.
\newblock {\em CERN-EP-2024-086, submitted to \PRL}, 2024.
\newblock \url{http://arxiv.org/abs/2403.15332}.

\bibitem{EXOT-2022-15}
{ATLAS Collaboration}.
\newblock {Search for light long-lived neutral particles from Higgs boson
  decays via vector-boson-fusion production from \(pp\) collisions at
  \(\sqrt{s} = 13\,\text{TeV}\) with the ATLAS detector}.
\newblock {\em CERN-EP-2023-226, submitted to {\em Eur. J. Phys. C}}, 2023.
\newblock \url{http://arxiv.org/abs/arXiv:2311.18298}.

\bibitem{CMS-PAS-EXO-23-013}
{CMS Collaboration}.
\newblock {Search for low-mass long-lived particles decaying to displaced jets
  in proton-proton collisions at $\sqrt{s} = 13.6~\mathrm{TeV}$}.
\newblock {\em CMS-PAS-EXO-23-013}, 2024.
\newblock \url{https://cds.cern.ch/record/2893044}.

\bibitem{CMS:2024gxp}
{CMS Collaboration}.
\newblock {Search for new physics with emerging jets in proton-proton
  collisions at $\sqrt{s}$ = 13 TeV}.
\newblock {\em CMS-EXO-22-015, submitted to JHEP}, 2024.
\newblock \url{http://arxiv.org/abs/arXiv:2403.01556}.

\bibitem{CMS-PAS-EXO-18-002}
{CMS Collaboration}.
\newblock {Search for heavy long-lived charged particles with large ionization
  energy loss in proton-proton collisions at $\sqrt{s} = 13~\mathrm{TeV}$}.
\newblock {\em CMS-PAS-EXO-18-002}, 2024.
\newblock \url{https://cds.cern.ch/record/2893595}.

\bibitem{EXOT-2019-06}
{ATLAS Collaboration}.
\newblock {Search for pair-production of vector-like quarks in lepton+jets
  final states containing at least one \(b\)-tagged jet using the Run~2 data
  from the ATLAS experiment}.
\newblock {\em CERN-EP-2023-254, submitted to {\em Phys. Lett. B}}, 2024.
\newblock \url{http://arxiv.org/abs/arXiv:2401.17165}.

\bibitem{CMS-B2G-20-014}
{CMS Collaboration}.
\newblock {A search for bottom-type vector-like quark pair production in
  dileptonic and fully hadronic final states in proton--proton collisions at
  \(\sqrt{s} = 13\,\text{TeV}\)}.
\newblock {\em CERN-EP-2024-016, submitted to {\em Phys. Rev. D}}, 2024.
\newblock \url{https://arxiv.org/abs/2402.13808}.

\end{thebibliography}

\end{document}